\documentclass[superscriptaddress,twocolumn,prb]{revtex4-1}

\usepackage{graphicx}

\usepackage{bm}
\usepackage{amsmath}
\usepackage{amssymb}
\usepackage{amsthm}

\begin{document}

\title{Aggregation and Electronically-Induced Migration of Oxygen Vacancies in TiO$_2$ Anatase}

\author{Martin Setvin}
\affiliation{Institute of Applied Physics, 
Vienna University of Technology, Wiedner Hauptstrasse 8-10/134, 1040 Vienna, Austria}

\author{Michael Schmid}
\affiliation{Institute of Applied Physics, 
Vienna University of Technology, Wiedner Hauptstrasse 8-10/134, 1040 Vienna, Austria}

\author{Ulrike Diebold}
\affiliation{Institute of Applied Physics, 
Vienna University of Technology, Wiedner Hauptstrasse 8-10/134, 1040 Vienna, Austria}


\begin{abstract} 
The influence of the electric field and electric current on the behavior of oxygen vacancies (V$_\mathrm{O}$s) in TiO$_2$ anatase was investigated with Scanning Tunneling Microscopy (STM). At the anatase~(101) surface V$_\mathrm{O}$s are not stable; they migrate into the bulk at temperatures above 200~K. Scanning a clean anatase~(101) surface at a sample bias greater than $\approx$+4.3~V results in surface V$_\mathrm{O}$s in the scanned area, suggesting that subsurface V$_\mathrm{O}$s  migrate back to the surface.  To test this hypothesis, surface V$_\mathrm{O}$s were first created through bombardment with energetic electrons. The sample was then mildly annealed, which caused the V$_\mathrm{O}$s to move to the subsurface region, where they formed vacancy clusters. These  V$_\mathrm{O}$ clusters have various, distinct shapes.   Scanning  V$_\mathrm{O}$ clusters with a high STM bias reproducibly converts them back into  groupings of  surface V$_\mathrm{O}$, with a configuration that is characteristic for each type of cluster. The dependence of the subsurface-to-surface V$_\mathrm{O}$ migration on the applied STM bias voltage,  tunneling current, and sample temperature was  investigated systematically. The results point towards a key role of energetic, 'hot' electrons in this process.   The findings are closely related to the memristive behavior of oxides and oxygen diffusion in solid-oxide membranes.

\end{abstract}

\maketitle

\section{Introduction}

The influence of electric fields on the behavior of oxygen vacancies (V$_\mathrm{O}$s) in metal oxides is of key importance for several applications of these materials. For example, the memristive switching in oxides is a promising approach for data storing.\cite{Strukov2008, Yang2013, Szot2014, Szot2011} While it is clear that field-induced redox reactions and field-induced material migration play a key role,\cite{Szot2011} very little is known about the detailed physical mechanisms and processes occurring at atomic scale.\cite{Zheng2012, Zheng2013} Similar phenomena are also essential in solid-oxide fuel cells, \cite{Minh1993, Steele2001} where oxygen is transported from the cathode to the anode through the  lattice of the oxide electrolyte (typically ZrO$_2$ or CeO$_2$) via migration of oxygen vacancies. 

Anatase is a metastable form of the prototypical metal oxide TiO$_2$,\cite{Diebold2003} a material that has been central in oxide research for decades. TiO$_2$ anatase is used in catalysis,\cite{Dohnalek2010} photocatalysis,\cite{Linsebigler1995, Henderson2011} and dye-sensitized solar cells.\cite{Gratzel2001} Owing to its promising electronical properties, it is also frequently investigated as a memristive material\cite{Szot2011, Szot2014} and as a Transparent Conductive Oxide (TCO).\cite{Furubayashi2005}  

Recently it was found that the anatase~(101) surface does not contain any surface V$_\mathrm{O}$s when prepared under ultrahigh vacuum (UHV) conditions by a standard sputter-annealing procedure.\cite{He2009} Surface V$_\mathrm{O}$s could be produced non-thermally by bombarding the sample with energetic electrons. These V$_\mathrm{O}$s proved unstable and diffused to the subsurface region at temperatures as low as 200~K.\cite{Scheiber2012} In our previous work\cite{Setvin2013Science} we have created the same atomic-size features by scanning a clean anatase~(101) surface at a high sample bias of $\approx$+5~V. We concluded that these were V$_\mathrm{O}$s that most likely originated from electronically induced migration of oxygen vacancies from the subsurface region to the surface.

In this work we investigate this phenomena in detail.  The paper is divided into three main parts.  In part  A we confirm that scanning a clean anatase~(101) surface at high positive sample biases creates surface V$_\mathrm{O}$s.  In part B we  verify that the V$_\mathrm{O}$s appearing on the surface indeed stem from the subsurface region.  To this end we performed the following experiment: First surface V$_\mathrm{O}$s were created by electron bombardment of a defect-free surface. Then the sample was annealed above room temperature. This resulted in the migration of V$_\mathrm{O}$s to the subsurface region, where they formed V$_\mathrm{O}$ clusters. Scanning at high bias voltage  reproducibly converted these clusters back into a collection of single surface V$_\mathrm{O}$s. This is a strong indication that the surface V$_\mathrm{O}$s form due to migration of material between the subsurface region and the surface and are not newly created by the STM tip. We also discuss the ramifications of the subsurface V$_\mathrm{O}$ clustering for surface reactivity.  In part C we systematically investigate how the rate of V$_\mathrm{O}$ creation depends on various parameters (bias voltage U$_\mathrm{S}$, tunneling current $I_\mathrm{T}$, temperature, and number of scans), and determine the role of the electric field $E$. Based on a quantitative analysis of the experimental results, we conclude that the V$_\mathrm{O}$ migration occurs close to the field-emission regime in STM, and it depends on $E$ only weakly under the experimental conditions used in this work. In part D we discuss the possible physical mechanisms underlying the STM-induced V$_\mathrm{O}$ migration. We identify the mechanism as one-electron process involving hot electrons.  The implications of our results for memristor research are discussed.
 
\section{Experimental details}

The experiments were performed in a UHV chamber at a base pressure below $1\times10^{-9}$~Pa, equipped with a commercial Omicron LT-STM head. STM measurements were performed at $T=78$~K or $T=6$~K. Electrochemically etched W STM tips were cleaned by Ar$^{+}$ sputtering and treated on a Au (110) surface to obtain a reproducible, metallic tip condition. An anatase mineral sample was cleaved\cite{Dulub2010} and cleaned \textit{in situ} by cycles of Ar$^{+}$ sputtering and annealing.\cite{Setvin2014SurfSci}  In  a  trace analysis\cite{Setvin2013Science}, the highest level of impurity in the sample was determined as 1.1 at.~\% Nb.   

Sample annealing was performed in a manipulator located in an adjacent preparation chamber (base pressure $3\times10^{-9}$~Pa).  The temperature was measured with a K-type thermocouple at the heating/annealing stage. We estimate that the quoted temperatures are accurate within $\pm10$~K. To create surface V$_\mathrm{O}$s, the sample was cooled to 105~K and irradiated with electrons from a thoroughly outgassed, rastered electron gun (500~eV, current density 8 $\mu$A cm$^{-2}$).

\section{Results and Discussion}
\subsection{STM-induced surface O vacancies}

\begin{figure}
    \begin{center}
        \includegraphics[width=1.0\columnwidth,clip=true]{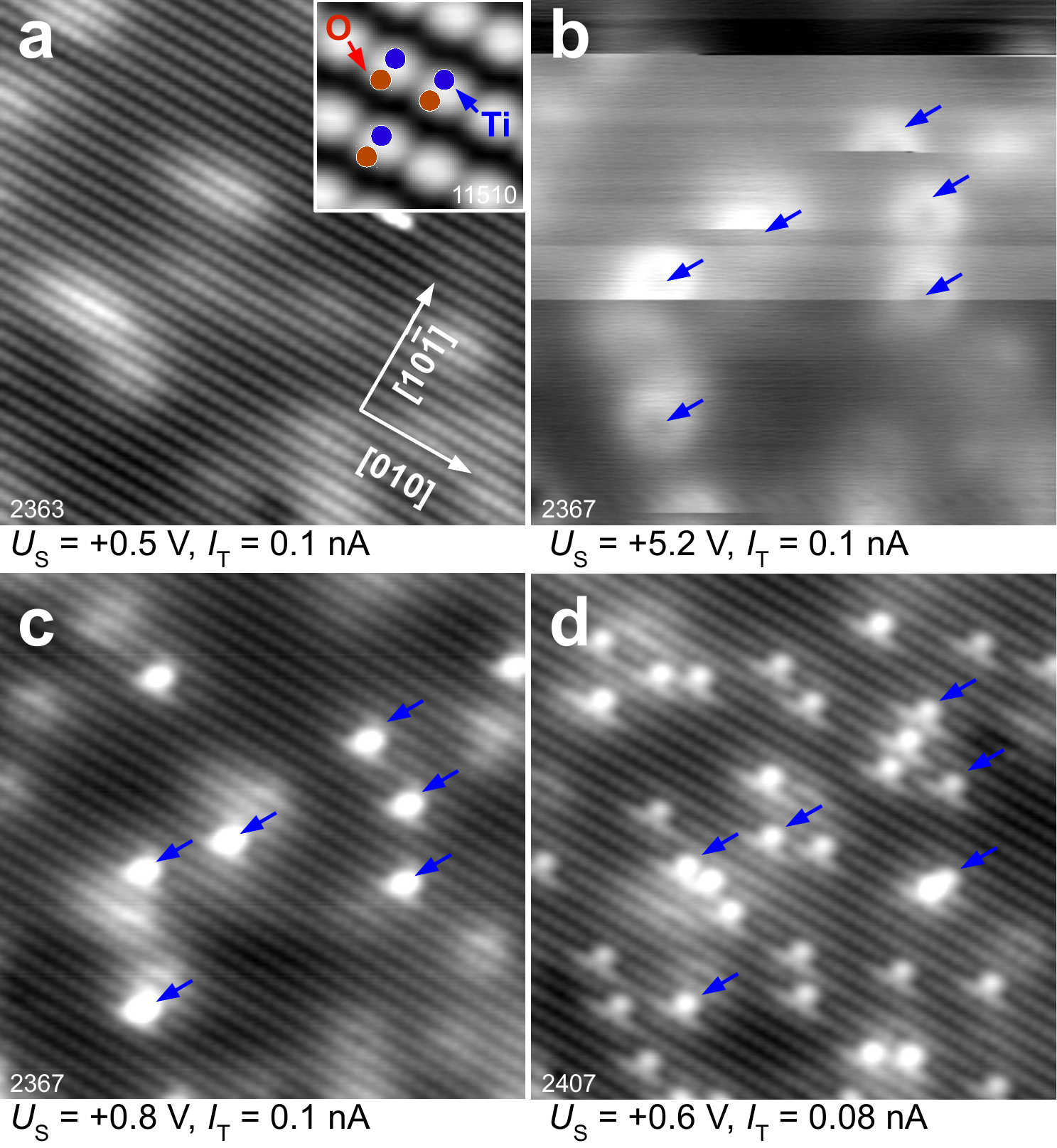}
    \end{center}
\caption{\label{Fig5} (a-d) Consecutive STM images (15$\times$15~nm$^2$, $T=6$~K) taken at the same area of an anatase~(101) surface. a) Clean surface. Inset shows a detailed STM image where the positions of surface 5-coordinated Ti and 2-coordinated O atoms are marked. b) STM image taken with a high bias ($U_\mathrm{Sample}$ = +5.2~V). c) Low-bias scan of the same area. Surface V$_\mathrm{O}$s that appeared during the high-bias scan are marked by blue arrows. d) Low-bias STM image after 32 high-bias scans  with conditions as in (b). The arrows point to the same V$_\mathrm{O}$s as in (c).}
\end{figure}

Figure~\ref{Fig5} shows the effect of scanning a clean anatase~(101) surface with STM at a high positive sample bias.  Single V$_\mathrm{O}$s appear within the scanned area (and, depending on the tip shape, the surrounding  few~nm, see ref.~\onlinecite{Setvin2013Science}).  Figure~\ref{Fig5}a displays the as-prepared, clean anatase~(101) surface; the STM image was taken with standard (low-bias) conditions. Brighter regions correspond to positions of subsurface donors,\cite{Ebert1999} likely extrinsic dopants. Fig.~\ref{Fig5}b shows the same area, but this time it is scanned at a high sample bias $U_\mathrm{S}=+5.2$~V and a tunneling current $I_\mathrm{T}=0.1$~nA. Horizontal streaks indicate structural changes that occurred during the high-voltage scan (marked by arrows). The same area is imaged again in Fig.~\ref{Fig5}c with normal imaging conditions (after two high-bias scans such as the one in (b)).  Several new features have appeared in the area; some are marked by arrows.  By comparison with $e$-beam induced defects\cite{Setvin2013Science, Scheiber2012} (see also Fig. ~\ref{Fig1} below)  these are identified as surface V$_\mathrm{O}$s.  When the same area is scanned multiple times at the elevated bias, more oxygen vacancies are created. This is  illustrated in Fig.~\ref{Fig5}d, which shows the same area after 32 high-bias scans at the tunneling parameters of Fig.~\ref{Fig5}b. For better orientation, the positions of the vacancies created in (b) are marked. It is rare that the V$_\mathrm{O}$s  move laterally by one position during a high-bias scan. We were not able to remove the  V$_\mathrm{O}$s with the STM tip, for example by applying a negative $U_\mathrm{S}$.


\begin{figure*}
    \begin{center}
        \includegraphics[width=2.0\columnwidth,clip=true]{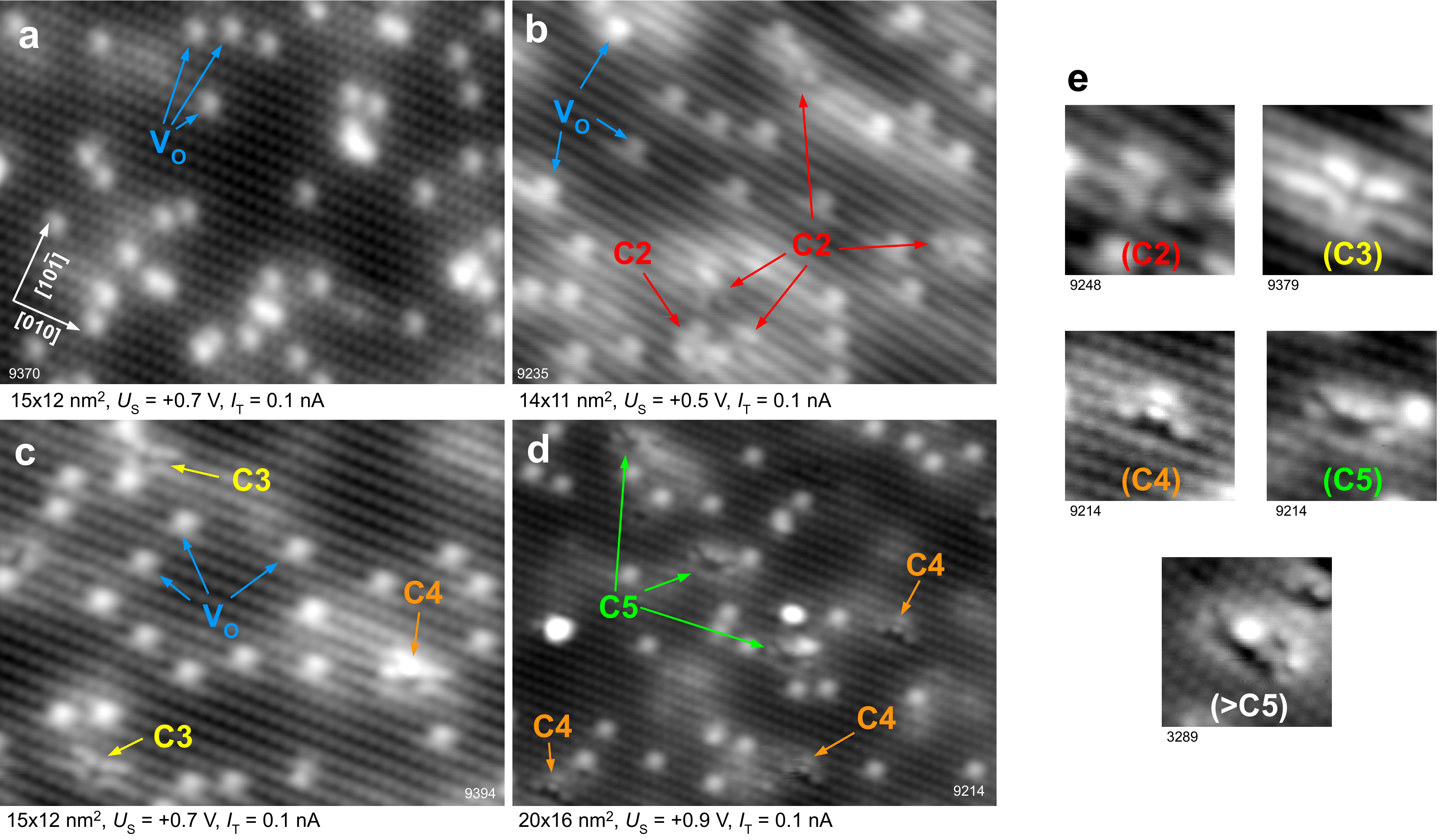}
    \end{center}
\caption{\label{Fig1} a) STM image of anatase (101) after exposing the surface to an electron beam at $T=105$~K (STM image taken at 6~K). Oxygen vacancies (marked V$_\mathrm{O}$) are generated. Electron bombarded-surfaces after annealing for 10 min to (b) 320~K, (c) 340~K, and (d) 380 ~K. Each panel represents a separate experiment and shows a different area on the sample. Some vacancies migrate to the subsurface and form characteristic V$_\mathrm{O}$ clusters, marked as C2-C5. (e) Details of the C2-C5 clusters.}
\end{figure*}


\subsection{Formation and Dissociation of V$_\mathrm{O}$ clusters}

\subsubsection{Subsurface vacancy clusters formed by aggregation} 

The surface V$_\mathrm{O}$s shown in the previous section could, in principle, be generated in two different ways: either by desorbing two-coordinated surface O$_\mathrm{2c}$ atoms (e.g., via electron or field-induced desorption), or by 'pulling' subsurface V$_\mathrm{O}$s to the surface. (More precisely, by pushing surface oxygen atoms deeper into the lattice.) Several previous observations point towards the second possibility.  First,  the threshold for electron-induced desorption is in the range of  tens of eV,\cite{Dulub2007} much higher than what can be achieved in STM.  Second, the tip is at a negative potential with respect to the sample, thus it should repel rather than attract the O anions. Third, on TiO$_\mathrm{2}$ rutile (110)  it was shown that the STM tip can move V$_\mathrm{O}$s laterally,\cite{Cui2008} so a vertical motion of V$_\mathrm{O}$s is not inconceivable.  Nevertheless we designed an experiment to test whether  it is indeed the exchange of O between subsurface and surface that gives rise to the effect shown in  Fig.~\ref{Fig5}.  We generated surface V$_\mathrm{O}$s (now by bombarding with electrons from a  conventional electron source)  and then annealed the surface. It is known that this results in the migration of the surface vacancies to the subsurface region.\cite{Scheiber2012} We then  located the subsurface V$_\mathrm{O}$s with STM, and pulled them back to the surface by applying a high sample bias.

Figure ~\ref{Fig1}a shows the anatase surface after exposing it to the electron beam. Surface V$_\mathrm{O}$s are marked. The V$_\mathrm{O}$ concentration decreases after annealing above 200~K, with no discernible trace in STM images.\cite{Scheiber2012}  However, when the annealing temperature exceeds room temperature, new features are observed.   Several types formed, with a characteristic and reproducible appearance in STM images. These are marked as C2-C5 in Figs.~\ref{Fig1}b-e. After annealing to temperatures slightly above 300~K, we observe mostly the features we call C2 (see Fig.~\ref{Fig1}c) and C3 (Fig.~\ref{Fig1}b). These two are imaged as distortions of the anatase~(101) lattice. They appear as protrusions, slightly shifted in the $[\bar{1}01]$ direction from the maxima of the bright rows. After annealing the surface to higher temperatures ($\approx$380 to 500~K), larger features appear, see C4 and C5 in Figs.~\ref{Fig1}b,d and also in ref.~\onlinecite{Scheiber2012}. An example of one feature larger than C5 is shown at the bottom of Fig.~\ref{Fig1}e. 

The features C2-C5 (as well as larger ones, which are not discussed here) are directly related to the oxygen deficiency in the near-surface region, as judged by the disappearance of surface V$_\mathrm{O}$s. We assume that they originate from clustering of the oxygen vacancies that were originally located on the surface, as discussed below. In the nomenclature C2-C5 the number denotes the order in which the clusters form, i.e., C2 forms at the lowest temperatures ($\approx$320~K) and so forth. The numbers 2-5 are  tentatively related to the number of oxygen vacancies in the cluster. The experiments shown below indicate that C2 contains two vacancies, C3 three. It is possible that these 'clusters' are nucleation centers of more reduced phases of titania, like Ti$_2$O$_3$ or TiO.


We note that the activation energy for hopping of a single surface V$_\mathrm{O}$ to the first subsurface layer is 0.75 eV according to DFT calculations.\cite{Scheiber2012, Cheng2009, Cheng2009JCP} A vacancy can possibly migrate deeper with lower activation barriers (as low as 0.17~eV). The energetically most favorable position for a single vacancy is likely in the first few subsurface layers rather than deep in the bulk; the switching between surface and subsurface sites that we observed for sample temperatures 200--300 K in our previous work\cite{Scheiber2012} supports the preferred residence of V$_\mathrm{O}$s in near-surface regions. In order to migrate in the direction parallel to the surface, the vacancy has to perform at least one hop with a calculated activation barrier of $\approx$1.1~eV.\cite{Cheng2009JCP} The difference in activation energies for vertical vs. lateral diffusion is consistent with our observation that the single surface V$_\mathrm{O}$s start to migrate subsurface at temperatures as low as 200~K, but the clusters start to appear above room temperature.

\subsubsection{Converting subsurface vacancy clusters into surface vacancies with the STM tip}

\begin{figure}
    \begin{center}
        \includegraphics[width=1.0\columnwidth,clip=true]{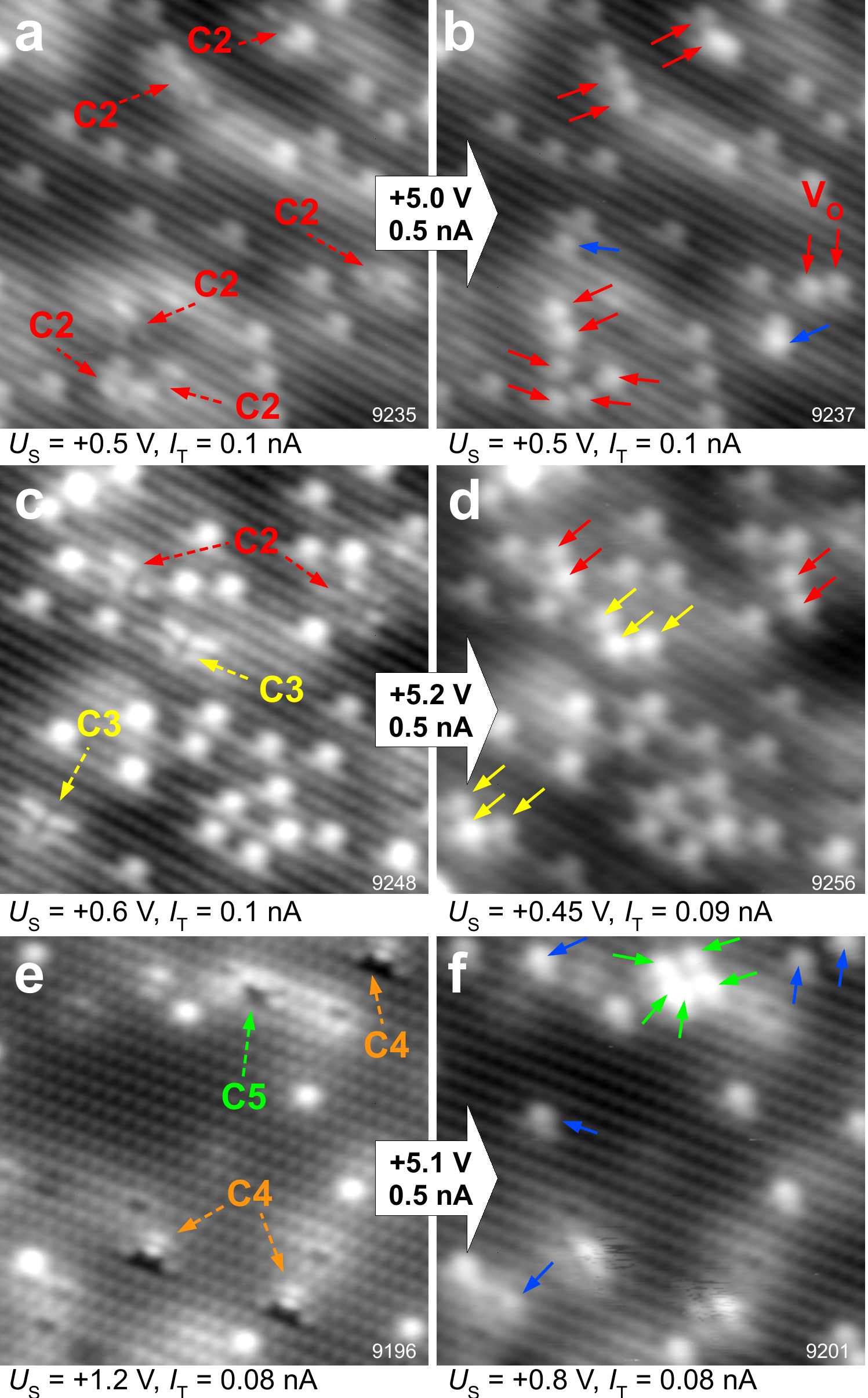}
    \end{center}
\caption{\label{Fig3} Electronically induced conversion of subsurface vacancy clusters back into single surface V$_\mathrm{O}$s. Panels (a,b), (c,d), and (d,e) show the same area before and after a high-bias scan (V$_\mathrm{S}\approx+5$~V, I$_\mathrm{T}=0.5$~nA). Dashed arrows in the left images show positions and types of vacancy clusters. Full arrows in the right images point to newly formed surface oxygen vacancies.}
\end{figure}

The V$_\mathrm{O}$ clusters C2-C5  were scanned with a high STM bias.  At a positive sample bias voltage of $\approx +5$~eV the clusters are converted  into  groupings of single V$_\mathrm{O}$s, with each one  characteristic for one type of cluster. This is shown in Fig.~\ref{Fig3}. The panels (a), (c), and (e) on the left show surfaces with various types of V$_\mathrm{O}$ clusters. These surfaces were prepared as described above, i.e., by bombarding a clean anatase (101) surface with electrons and subsequent annealing to 320, 320, and 380~K, respectively. The STM images in panels (b), (d), and (f) at the right were taken after each area was scanned at the bias voltages and high tunneling currents  indicated in the large arrows in Fig.~\ref{Fig3}. 

The high-bias scans cause the disappearance of the surface distortions that are indicative of the V$_\mathrm{O}$ clusters.  The original $(1\times 1)$ surface is obtained and surface V$_\mathrm{O}$s appear, with a number that is characteristic for each cluster. The cluster C3 is always converted into three V$_\mathrm{O}$s that are usually arranged in the  triangular pattern shown in Fig.~\ref{Fig3}d. Cluster C2 is converted into a pair of oxygen vacancies. Larger clusters (C4 and higher) can rarely be converted into single oxygen vacancies (see Fig.~\ref{Fig3}e, f), even if higher electric fields are applied. Usually the cluster undergoes certain changes and some surface V$_\mathrm{O}$s appear, but the original (1$\times$1) surface could not be restored. Probably the crystal lattice is too distorted for the larger clusters; in addition, interstitial Ti atoms\cite{Wendt2008} may already be involved.

The high-bias scans above the clusters were performed both in the constant current and the constant height modes, providing the same result. (We performed this test to ensure that the tip-sample distance is the same above the cluster and the clean surface.) Compared to the sample bias needed to obtain a comparable number of V$_\mathrm{O}$s on a clean surface (the experiment in Fig.~\ref{Fig5}), the bias for converting subsurface V$_\mathrm{O}$ clusters to surface V$_\mathrm{O}$s is only $\approx 0.3$\,V lower. The similar values indicate that the physical mechanism is the same in both cases: The surface V$_\mathrm{O}$s are not generated by desorbing oxygen atoms from the surface, but by transport of material within the sample.
 
\begin{figure}
    \begin{center}
        \includegraphics[width=0.8\columnwidth,clip=true]{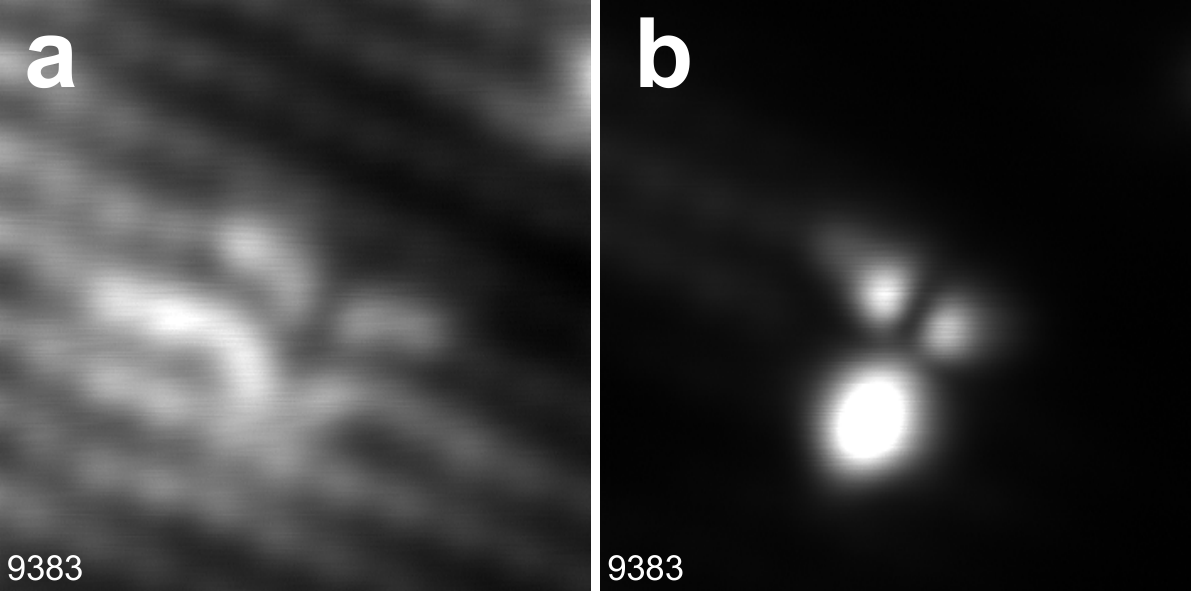}
    \end{center}
\caption{\label{Fig4} Empty- and filled-state STM image of a C3 cluster. Constant height images measured at the same position and the same tip--sample distance. a) $U_\mathrm{S}=+0.5$~V, b)~$U_\mathrm{S}=-0.6$~V.}
\end{figure}

We have investigated the electronic structure of the V$_\mathrm{O}$ clusters. All the clusters show localized states below the Fermi level, an indication of trapped electrons.\cite{SetvinPolarons}  As an example, we show empty and filled-states STM images of a C3 cluster in Figs.~\ref{Fig4}a,b, respectively. The filled-state image of the C3 cluster shows one large and two small spots. The surrounding anatase lattice shows negligible LDOS below $E_\mathrm{F}$, as expected.\cite{SetvinPolarons} Scanning tunneling spectroscopy measurements of the C3 cluster (data not shown here) show  states at $\approx -0.3$ (larger spot) and $\approx -0.6$~eV (two smaller spots) below the Fermi level. These states appear shallower than the state for a single surface V$_\mathrm{O}$, which is typically $-1.0$~eV below E$_\mathrm{F}$.\cite{Setvin2014Angewandte, SetvinPolarons}

 \subsubsection{Impact of the V$_\mathrm{O}$ clustering on the materials properties}

Our results clearly show that V$_\mathrm{O}$s tend to form clusters in the subsurface region. In TiO$_2$ rutile, the extended defects form already at V$_\mathrm{O}$ concentrations as low as 0.001.\cite{Bursill1984Nature, Bursill1984JSSCH} No such reports exist for anatase so far, but existence of the VO clusters in reduced anatase also seems consistent with photoelectron spectroscopy data, where a considerable Ti$^{2+}$ signal was detected in a synchrotron-beam-damaged material\cite{Jackman2015} (compared to usual Ti$^{4+}$ and Ti$^{3+}$ in rutile).\cite{Diebold2003} The V$_\mathrm{O}$ clustering  must be due to a lower energy as compared to single V$_\mathrm{O}$s. As charged vacancies will repel each other, this indicates that V$_\mathrm{O}$ clusters are neutral or carry only a single negative charge per cluster. We note that the vacancy charge state can be possibly influenced by other defects nearby, thus the clustering may be affected by impurities in the material for example. 

The presence of the V$_\mathrm{O}$ clusters needs to be considered in the various applications of this materials.  For example, single subsurface V$_\mathrm{O}$s are frequently considered in calculations of chemical reactions. In previous studies it was proposed that subsurface V$_\mathrm{O}$s can migrate back to the surface upon adsorption of certain species, and directly participate in chemical reactions.\cite{Setvin2013Science, Li2014} It was  predicted theoretically that adsorption of an O$_2$ molecule above a subsurface V$_\mathrm{O}$ should result in the migration of the V$_\mathrm{O}$ towards the surface and  in a bridging interstitial dimer (O$_2$)$_\mathrm{O}$.\cite{Setvin2013Science} However, in the same work it was shown experimentally that this reaction occurs rarely. The number of (O$_2$)$_\mathrm{O}$ features obtained  in this way was only on the order of 0.1 to 1\%~ML. A similar reaction was theoretically predicted for H$_2$O adsorption. Reaction of a H$_2$O molecule with a subsurface V$_\mathrm{O}$ should result in two bridging hydroxyl groups.\cite{Li2014} This reaction has never been observed experimentally, however.  Instead it is known that water adsorbs molecularly on the anatase~(101) surface.\cite{Aschauer2010Water} The V$_\mathrm{O}$ clustering observed here may explain why it is rare that the vacancies enter the chemical reactions. Theoretical works so far only investigated the configuration of a single subsurface oxygen vacancy. Our results indicate that more favorable configurations exist, where several V$_\mathrm{O}$s are bunched together in a cluster. Breaking such cluster is therefore energetically more costly than moving a single V$_\mathrm{O}$ from the subsurface region to the surface. The single subsurface V$_\mathrm{O}$s are probably a rare species on the anatase~(101) surface compared to $e. g.$ oxygen vacancies on the rutile~(110) surface.  

The observed behavior of the V$_\mathrm{O}$ clusters is also closely related to the memristor research. We have shown that annealing of a reduced anatase surface leads to the clustering of the V$_\mathrm{O}$s. On the other hand, the clusters can be disassembled back into single V$_\mathrm{O}$s under influence of applied electric field and tunneling current. It is known that the formation of extended defects and reduced Magneli phases\cite{Bursill1984Nature, Bursill1984JSSCH} and their transformation back into the stoichiometric oxide is a key process in the memristive switching. We have observed the initial step of the V$_\mathrm{O}$ aggregation and decomposition. In the following section, we perform a statistical analysis of the experimental data and characterize physical mechanisms responsible for the V$_\mathrm{O}$ migration.

\subsection{Physical mechanisms of STM-induced V$_\mathrm{O}$ migration}

\subsubsection{Dependence of V$_\mathrm{O}$ formation on STM parameters}

\begin{figure}
    \begin{center}
        \includegraphics[width=1.0\columnwidth,clip=true]{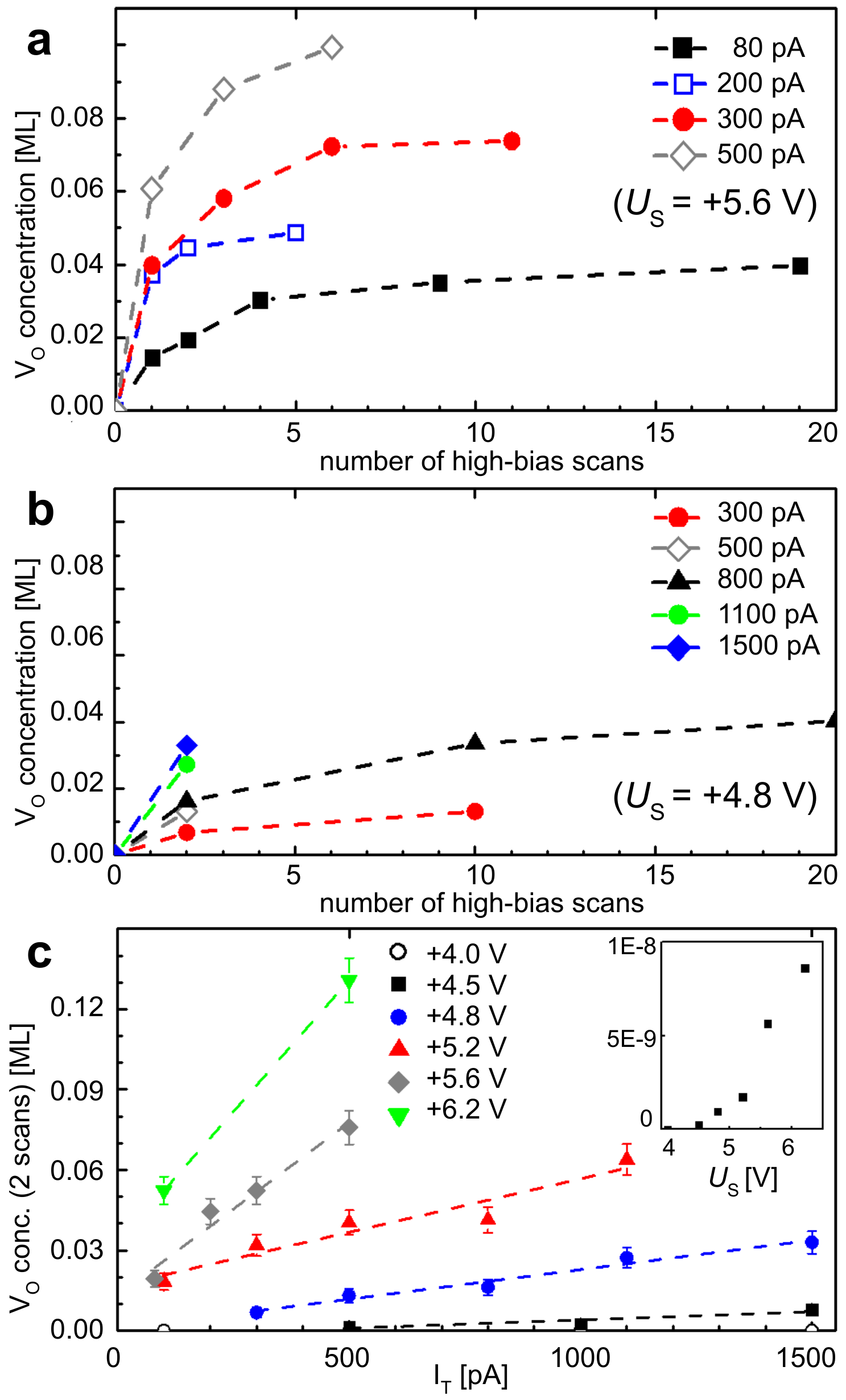}
    \end{center}
\caption{\label{Fig6} (Color online) a) Concentration of V$_\mathrm{O}$s after scanning multiple times at a high  STM bias (such as in Figure~\ref{Fig5}). Each scan was conducted at $U_\mathrm{S}=+5.6$~V and the indicated tunneling currents. b) The same for a lower $U_\mathrm{S}=+4.8$~V. c) Concentration of V$_\mathrm{O}$s obtained in two high-bias scans, plotted as a function of $U_\mathrm{S}$ and $I_\mathrm{T}$. The inset shows the slopes of the curves as a function of U$_\mathrm{S}$. All data obtained at $T=78$~K with the same STM tip.}
\end{figure}

For simplicity, we focus on single vacancies, i.e., we start each experiment with a clean surface and conduct experiments as the one laid out in Figure~\ref{Fig5}.   We investigate how the sample bias, tunneling current, and number of scans affect the concentration of V$_\mathrm{O}$s that form during such high-bias scans. Our main findings are summarized in Fig.~\ref{Fig6}. In Fig.~\ref{Fig6}a, we chose a fixed $U_\mathrm{S} = +5.6$~V for the high-bias scans and investigated the concentration of V$_\mathrm{O}$s in dependence of the number of the scans. This procedure was repeated for different tunneling currents. A new surface area was used for each value of $I_T$.  During all experiments shown in Fig.~\ref{Fig6} the size of the scan area and the total scan time were kept constant at  $15\times 15$~nm$^2$ and $\tau = 100$~s, respectively. A similar data set with a lower $U_\mathrm{S} = +4.8$~V is shown in Fig.~\ref{Fig6}b.

The results in Fig.~\ref{Fig6}a show that the V$_\mathrm{O}$ density approaches a saturation value that depends on the tunneling parameters. We then investigated the rate of V$_\mathrm{O}$ generation in the zero V$_\mathrm{O}$ concentration limit, $i.e.,$ the initial slope of the curves in Fig.~\ref{Fig6}a. This is depicted in Fig.~\ref{Fig6}c for different combinations of $U_\mathrm{S}$ and $I_\mathrm{T}$. Plotted is the V$_\mathrm{O}$ concentration obtained after the first two high-bias scans.  Within the range of currents investigated here (80 to 1500~pA) this initial rate is roughly proportional to the tunneling current, a clear indication for a single-electron process. The proportionality is less precise in the data sets measured at higher $U_\mathrm{S}$, as the $n(t)$ curves approach the saturation value faster and the assumption of the zero V$_\mathrm{O}$ concentration is not  perfectly fulfilled.   

To estimate the dependence on the sample bias, the curves in Fig.~\ref{Fig6}c were fitted by a linear function $y=a I_\mathrm{T}+b$. The slope $a$ is plotted in the inset of Fig.~\ref{Fig6}c. Here the $y$-axis is given in units of cross section $\sigma=a /(\tau I_\mathrm{T} A)$, using the known value of the scanned area $A$ and scan time per frame $\tau$. No vacancies are generated below a threshold bias of $U_\mathrm{S} = +4.3$~V. From the available data we judge that the dependence follows a polynomial behavior above the threshold, likely quadratic or cubic. 

The data shown in Fig.~\ref{Fig6} have been measured at temperature of 78~K. A similar experiment was performed at $T=6$~K, with a comparable result. This suggests that the V$_\mathrm{O}$ migration is not activated thermally.

\subsubsection{Electric field}

The results  in Fig.~\ref{Fig6} clearly show that a threshold $U_\textbf{S}$ (or perhaps a threshold electric field $E$) is necessary  so that subsurface V$_\mathrm{O}$s move to the surface (or, conversely, surface O are pushed into the lattice)  during STM scans. The initial rate of surface/subsurface O exchange is rougly proportional to $I_\mathrm{T}$, suggesting that a one-electron process is involved.  The rate shows a polynomial dependence on $U_\mathrm{S}$ above the threshold of +4.3~V.  The tip-induced vacancy migration is self-limiting, and the final V$_\mathrm{O}$ concentration after many scans also depends on $U_\textbf{S}$ and $I_\mathrm{T}$ (Fig.~\ref{Fig6}c).

In tip-induced processes, the tunneling current $I_\mathrm{T}$, bias voltage $U_\textbf{S}$, and electrical field $E$  inside the sample\cite{Selcuk2014} can all play a role. It is not trivial to disentangle how $E$ depends on the tunneling parameters.  The tip-sample distance $d$ depends on $U_\textbf{S}$ and $I_\mathrm{T}$, and  only part of the field penetrates into the semiconductor due to screening.  This 'tip-induced band-bending (TIBB)', in turn, can depend on the presence of surface V$_\mathrm{O}$s.

\begin{figure}
    \begin{center}
        \includegraphics[width=1.0\columnwidth,clip=true]{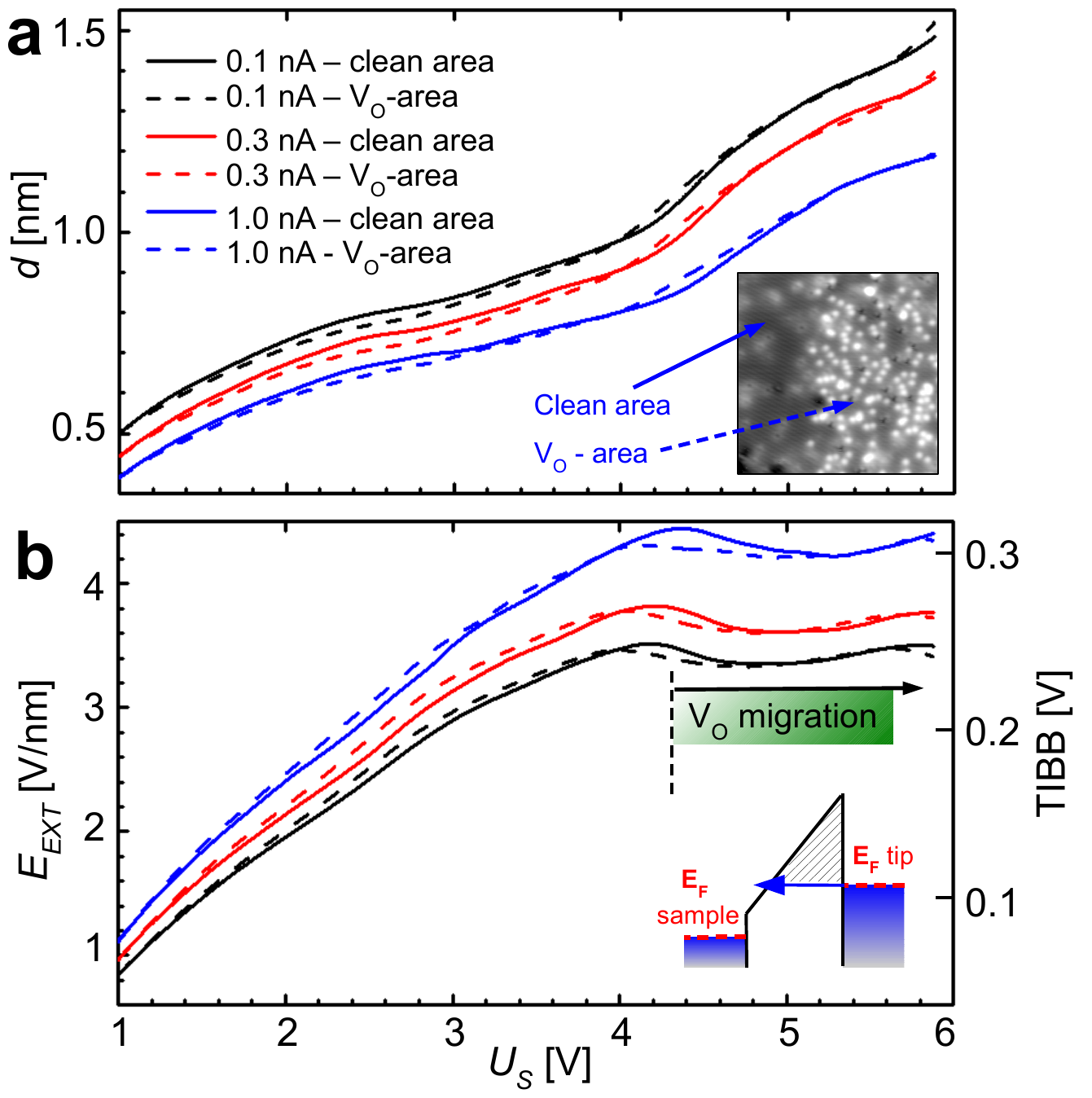}
    \end{center}
\caption{\label{Fig7} (Color online) Estimation of the electric field between the tip and the sample. a) $d(U_\mathrm{S})$ curves measured above the surface with closed feedback loop at different tunneling currents. The curves were measured above the clean surface and in an area with surface V$_\mathrm{O}$s as shown. Measured at $T=6$~K. $d$ is the absolute tip-sample distance. b) Calculated electric field between the tip and the sample. The right axis shows an estimate of the tip-induced band-bending (TIBB) inside the sample. The tunneling scheme in the inset explains why the field only depends on the $I_\mathrm{T}$ at higher $U_\mathrm{S}$ values (see the text for details).}
\end{figure}

To determine  the electric field at the conditions where V$_\mathrm{O}$s appear, we performed local spectroscopy measurements. To avoid  high tunneling currents,  $z(U_\mathrm{S})$ spectroscopy with a closed feedback loop was used. The measurement time for the curves was set as low as possible (100 ms/curve), in order to avoid additional creation of vacancies during the measurements. The data shown are averages of 10 curves measured at different positions. Figure~\ref{Fig7}a shows the result in terms of the absolute tip-sample separation $d$.   We calibrated $d$ by 'touching the surface' (see refs.~\onlinecite{Jelinek2008, SetvinPolarons}).  For $U_\mathrm{S}=+1.0$~V and $I_\mathrm{T}=0.1$~nA, we estimate that $d_0=0.5$~nm, all other values can be derived by measuring $z(U_\mathrm{S})$ and $z(I)$ spectra.  

In Fig.~\ref{Fig7}b we converted the measured $z(U_\mathrm{S})$ curves into the external electric field acting between the tip and the surface. The electric field is approximately $E_\mathrm{EXT}=U_\mathrm{S}/d$.  A more precise expression
 
\begin{equation}
E_\mathrm{EXT}=\frac{0.94(U_\mathrm{S}-U_\mathrm{LCPD})}{d}
\end{equation} 

takes into account the local contact potential difference between the tip and the sample, $U_\mathrm{LCPD}$. We use a value of +0.6~V, assuming work functions of 5.1 and 4.5~eV for the anatase surface and the tip, respectively.\cite{Xiang2007, Scanlon2013, LothThesis}   
The factor 0.94 in eqn. (1) is a correction for the TIBB -- we estimate that $4-8\%$ of the applied bias penetrates into the sample. This value was obtained with Feenstra's Poisson equation solver,\cite{Feenstra2005} for various combinations of relevant input parameters. This relatively low level of TIBB results from the high dielectric constant of anatase ($\epsilon _r \approx 35$),\cite{Selcuk2014, Berger1997} and high sample doping. (Our  sample is Nb doped\cite{Setvin2013Science} with $n=10^{20}$~to~$10^{21}$~cm$^{-3}$.)

In order to know whether the surface V$_\mathrm{O}$s influence the electric fields acting in the system, we performed $z(U_\mathrm{S})$ measurements  in a region where such V$_\mathrm{O}$s had been created with the STM tip, see dashed lines in Fig.~\ref{Fig7}.  The spectra were taken at  least two lattice constants away from V$_\mathrm{O}$s (concentration of 7\%).  There is a small, but reproducible difference in $z(U_\mathrm{S})$ measurements and  $E_\mathrm{EXT}(U_\mathrm{S})$ estimates between clean and defective areas.

The curves in Fig.~\ref{Fig7}b show a plateau above $U_S=+4$~V. This is explained by the tunneling scheme in the inset of Fig.~\ref{Fig7}b. When the applied $U_\mathrm{S}$ approaches 
the sample work function, the tunneling barrier (hatched area) only depends on the electric field between the tip and the sample. The field emission regime is almost reached.  Interestingly, the onset of this plateau coincides with the threshold   for tip-induced vacancy creation, $U_\mathrm{S} = +4.3$~V. The electric field is roughly constant in the whole regime of tip-induced vacancy migration ($U_\mathrm{S} \geq 4.3$\,V; inset in Fig. \ref{Fig7}b). On the other hand, the rate of V$_O$ migration increases dramatically with $U_\mathrm{S}$ in this range. This clearly shows that the voltage dependence is not related to the field. The values of the electric field in the regime of vacancy migration, 3.3--4.5\,V/nm (depending on $I_\mathrm{T}$, see Fig.~\ref{Fig7}b) correspond to a TIBB of 0.25--0.30\,V inside the sample.


This allows direct comparison of our experimental conditions to the density functional theory calculations performed by Selcuk \textit{et. al}.\cite{Selcuk2014} The calculations show that an electric field inside the anatase sample affects the energetics of surface $vs.$ subsurface V$_\mathrm{O}$s. The surface V$_\mathrm{O}$ should become $\approx$0.2~eV more favorable at electric fields comparable to our experimental conditions. On the other hand, the field does not reduce the energy barrier for V$_\mathrm{O}$ migration from the subsurface to surface, which remains 0.5 to 0.9~eV. In the following we will argue that hot electrons that are injected from the STM tip into the sample help overcoming this activation barrier.

\subsubsection{Quantitative analysis}

We have measured data sets similar to the one in Fig.~\ref{Fig6}a for five different values of $U_\mathrm{S}$. All data shown below were obtained with a single STM tip. Here we analyze these results to determine how the V$_\mathrm{O}$ concentration scales with the physical quantities $E$, $U_\mathrm{S}$, $I_\mathrm{T}$, and the time $t$. 

\begin{figure}
    \begin{center}
        \includegraphics[width=1.0\columnwidth,clip=true]{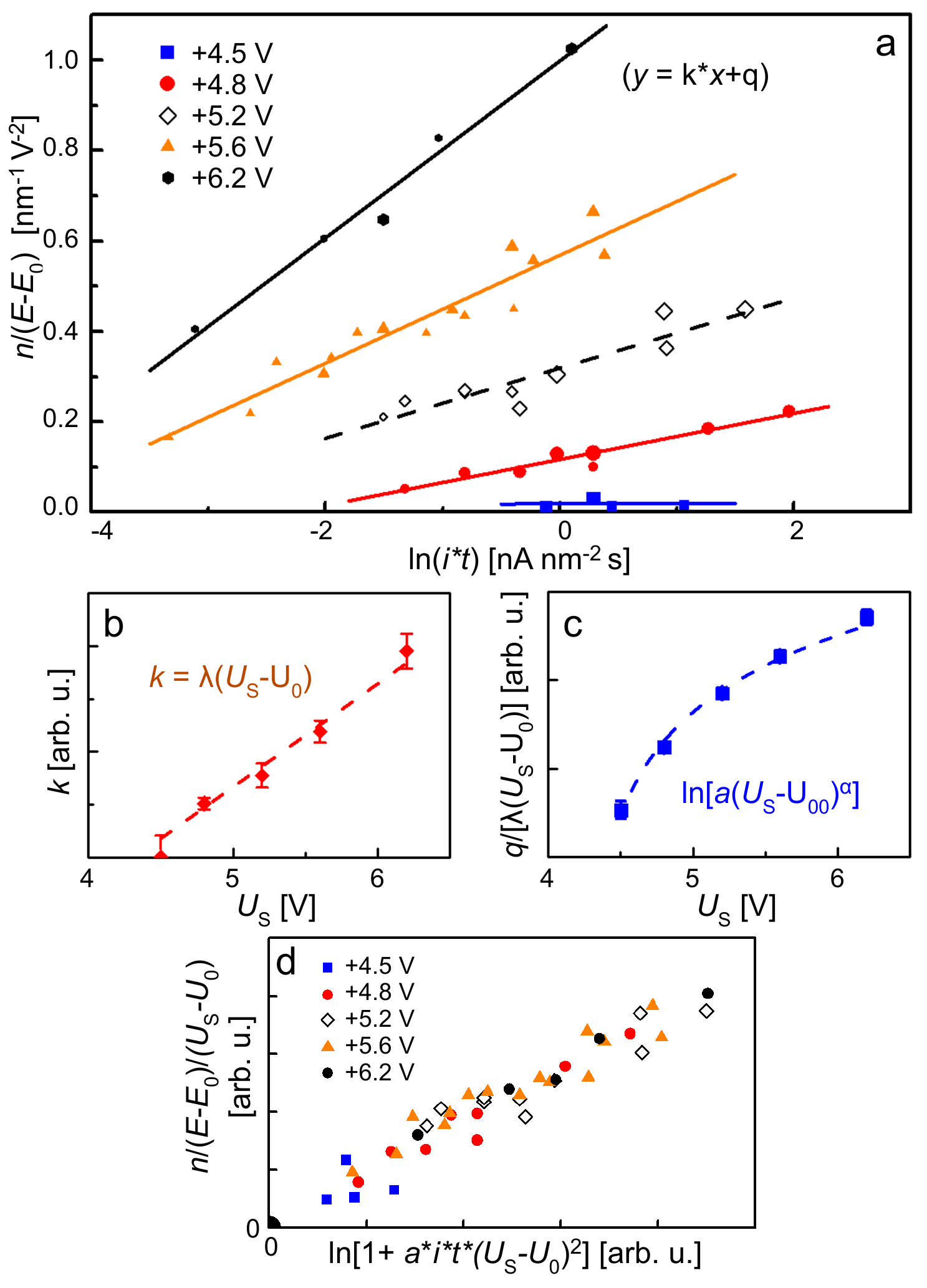}
    \end{center}
\caption{\label{Fig10} (Color online) a) Experimentally measured V$_\mathrm{O}$ concentration as a function of the tunneling current $I_\mathrm{T}$ (normalized to the scan area, $i=I_\mathrm{T}/A$), sample bias $U_\mathrm{S}$, and scanning time $t$. The size of the data points reflects the tunneling current used ($I_\mathrm{T}$, which ranged from 0.08 to 1.5~nA). See text for scaling of the axes.  The data set for each $U_\mathrm{S}$ was fitted by a linear function $y=kx+q$. The fitting parameters $k$ and $q$ are further analyzed in b) and c) as a function of $U_\mathrm{S}$ (see also Appendix). d) with a suitable scaling, all experimental data points collapse to a single line.} 
\end{figure}

The data are plotted in Fig.~\ref{Fig10}. The points measured for each single value of $U_\mathrm{S}$ collapse to a single line if we scale the axes in a suitable way. Our initial data analysis has shown that the V$_\mathrm{O}$ migration is initiated by a single-electron process. Thus an appropriate scale for the $x$-axis is the time multiplied by the current density $i$. This product is the electron dose per unit area applied during a scan or a sequence of scans; it does not depend on how the tunneling current is distributed under the tip. The $y$-axis shows the V$_\mathrm{O}$ concentration $n$, divided by the electric field $(E-E_0)$. We found that this scaling is necessary to collapse the values for different $I_\mathrm{T}$ on one line. The value of $E$ for each data point was determined from Fig.~\ref{Fig7}b. The scaling used in Fig.~\ref{Fig10}a efficiently separates the effects of $I_\mathrm{T}$ and $E$, as all the data sets obtained for single values of $U_\mathrm{S}$ follow lines on a linear-log scale. The parameter $E_0$ was varied to maximize the R-factors of linear fits in all data, resulting in $E_0 = 2.6 \pm 0.4$~V/nm. We note that scaling the $y$-axis by $(E-E_0)$ is a simple approximation of any $E$-dependence, as the range of $E$ used in our experiments is very small (Fig.~\ref{Fig7}b). Scaling by $E^2$ or $E^3$ provides a very similar result.


The data points in Fig.~\ref{Fig10}a are fitted by linear curves 

\begin{equation}
\label{EqFitting}
\frac{n}{(E-E_0)} = k \cdot \mathrm{ln} (it) + q,
\end{equation} 

where $k$ and $q$ now only depend on $U_\mathrm{S}$. By fitting $k(U_\mathrm{S})$ and $q(U_\mathrm{S})$ by suitable functions (shown in Fig.~\ref{Fig10}b,c) and inserting these functions into Eq.~\ref{EqFitting}, we arrive at an analytical expression 

\begin{equation}
\label{EqFitting3}
n = \lambda(E-E_0)(U_\mathrm{S}-U_0) \cdot \mathrm{ln} [a (U_\mathrm{S}-U_0)^2 i t + 1]. 
\end{equation} 

Here $U_0 = 4.3 \pm 0.3$~V and $a$ and $\lambda$ are constants. (For details of the fitting, including the meaning of parameters displayed in  Fig.~\ref{Fig10}c and the origin of the  "+1" in  Eq.~\ref{EqFitting3}, see the Appendix.) Using Eq.~\ref{EqFitting3} we can scale all experimental data points into a single linear dependence, see Fig.~\ref{Fig10}d. We note that Eq.~\ref{EqFitting3} is the solution of equation

\begin{equation}
\label{EqDerivation}
\frac{\mathrm{d}n}{\mathrm{d}t} = a i (U_\mathrm{S}-U_0)^{3} (E-E_0) \mathrm{exp}(-\frac{n}{\lambda (E-E_0) (U_\mathrm{S}-U_0)}),
\end{equation} 

where $\mathrm{d}n/\mathrm{d}t$ is the vacancy flow towards the surface. In the following we analyse Eq.\ref{EqDerivation} to gain insight into the physics of the STM-induced vacancy migration.

\subsection{Discussion of possible physical mechanisms}

The equation has a form of d$n$/d$t \approx R \mathrm{e}^{-\alpha n}$. Here $R$ is a rate, which is exponentially "damped" by V$_\mathrm{O}$s already present at the surface. $R$ is proportional to the current density $i$. The term $(U_\mathrm{S}-U_0)^{3}$ has the largest influence on the cross-section of the process.  The STM bias voltage $U_s$ is the maximum energy of the electrons injected into the sample.  A minimum energy ($U_0$) is needed, and the rate strongly increases with the electron energy above this threshold. In other words, "hot electrons" are needed. One possibility is that the electrons must be injected into a specific electron state in the conduction band to initiate the V$_\mathrm{O}$ migration. The importance of injecting the electrons into $s$-orbitals has previously been proposed for tip-induced migration of hydrogen atoms absorbed in bulk palladium.\cite{Rey2012, Sykes2005, Mitsui2007} While such a process must remain speculative at this point, we note that the experimental value of $U_0 = +4.3$~V matches the region where the $d$-character of the conduction band changes to $s$-like according to theoretical calculations of the anatase band structure.\cite{Landmann2012, SetvinPolarons} 

The damping term $\mathrm{exp} \left\{-n/[\lambda(E-E_0)(U_\mathrm{S}-U_0)]\right\}$ indicates that the efficiency of the electronic excitations decreases with growing surface V$_\mathrm{O}$ concentration. The influence of the V$_\mathrm{O}$s already present on the surface can be suppressed by applying a higher electric field and a higher $U_\mathrm{S}$. The first term could be due to surface V$_\mathrm{O}$s screening the field penetrating into the sample.  This decreases the TIBB, which is a necessary component of the V$_\mathrm{O}$ migration process. The dependence on $U_\mathrm{S}$ may be related to scattering the hot electrons at the surface V$_\mathrm{O}$s. Each V$_\mathrm{O}$ provides two localized electrons with a state $\approx$1~eV deep in the band gap. The electrons injected from the tip can possibly excite the V$_\mathrm{O}$-electrons to the conduction band, resulting in a significant energy loss of the primary electron. The $(U_\mathrm{S}-U_0)$ term in the exponential would be then related to the cross-section of this electron-electron scattering process.

A brief summary of the role of $U_\mathrm{S}$ and $I_\mathrm{T}$ is following: $U_\mathrm{S}$ determines the energy of electrons injected into the material. This in turn determines to which particular energy band in the conduction band they are injected and the energy available for single-electron processes. The tunneling current influences the process in two ways. First, it determines the rate of electronic excitations (rate $R$ is a linear function of $I_\mathrm{T}$). Second, $I_\mathrm{T}$ is linked to the electric field acting in the system (see Fig. \ref{Fig7}). We note that for constant-current conditions, the electric field does not change significantly with $U_\mathrm{S}$ in the regime used for inducing the V$_\mathrm{O}$ migration.

Our finding that hot electrons play a role in the V$_\mathrm{O}$ migration is in agreement with memristor research. In ref.~\onlinecite{Szot2014} it is argued that simple Joule heating may not be the only mechanism involved in defect migration within the crystal. Hot electrons may be directly scattered at defects in the oxide lattice, providing energy for material transport. Our experiments also show that the electric field layout inside the sample plays a certain role. The field helps to revert the energy balance in the material, providing a direction for the vacancy flow. The value of the electric field varies only slightly under our experimental conditions, therefore the field-dependence could not be exactly characterized and was only approximated by a linear function in the equations.

\section{Conclusions}

We have shown that scanning the anatase~(101) surface at high positive sample bias results in the appearance of surface V$_\mathrm{O}$s in the scanned area. We attribute this effect to a migration of V$_\mathrm{O}$s from the subsurface region to the surface. The process is self-limiting: presence of V$_\mathrm{O}$s on the surface prevents further subsurface-to-surface V$_\mathrm{O}$ migration. Analysis of the experimental data indicates that the electric field penetrating into the sample is an important factor for reverting the  energy balance between the surface and subsurface V$_\mathrm{O}$s. 
The hot electrons injected from the tip provide the activation energy necessary for the V$_\mathrm{O}$ migration through the lattice.  

It was further shown that V$_\mathrm{O}$s can easily form subsurface clusters upon annealing. We identified V$_\mathrm{O}$ clusters that contain two to five vacancies.  Likely this is the initial step in the  formation of extended defects and reduced TiO$_{2-x}$ phases. Subsurface aggregates of  V$_\mathrm{O}$s can be converted back into single surface V$_\mathrm{O}$s by applying a suitable electric field. This process closely resembles  memristive switching: Two distinct states exist, one that is reached upon thermal annealing and another one by applying a high electric field. The memristive behavior of oxides have been investigated for more than 50 years, yet there is essentially no knowledge about processes occurring at atomic scale. Our results could provide a significant step forward to identifying the underlying physical mechanisms. 


\begin{acknowledgments}

{This work was supported by the ERC Advanced Research Grant `OxideSurfaces',
and by the Austrian Science Fund (FWF) under project number F45.}

\end{acknowledgments}

 
%

\appendix*

\section {Fitting details}

Fitting the data points in Fig.~\ref{Fig10}a by $y=kx+q$ provides dependencies $k(U_\mathrm{S})$ and $q(U_\mathrm{S})$ displayed in Figs.~\ref{Fig10}b,c, respectively. The $k(U_\mathrm{S})$ appears linear, $\mathrm{k}=\lambda (U_\mathrm{S}-U_0)$, with $U_0 = 4.3 \pm 0.4$~V and $\lambda = 0.10 \pm 0.01$~V$^{-2}$nm$^{-1}$. One can rewrite Eq.~\ref{EqFitting} as

\begin{equation}
\label{EqFitting2}
\frac{n}{\lambda(E-E_0)(U_\mathrm{S}-U_0)} = \mathrm{ln} (it) + \frac{\mathrm{q}}{\lambda(U_\mathrm{S}-U_0)}.
\end{equation} 

We can fit the quantity $\mathrm{q}/[\lambda(U_\mathrm{S}-U_0)]$ as a function of $U_\mathrm{S}$ (see Fig.~\ref{Fig10}c). A suitable function is logarithmic, we use an ansatz of $\mathrm{ln}[a(U_\mathrm{S}-U_{00})^\alpha]$.  Based on the initial analysis in Fig.~\ref{Fig6} we guess an  $\alpha$ close to 2, and $U_{00}$ similar to $U_0$. Fitting the plot in Fig.~\ref{Fig10}c indeed provides $\alpha = 2.2 \pm 0.5$, $U_{00} = 4.2 \pm 0.2$~V, and $\mathrm{a} = 40 \pm 20$~A$^{-1}$s$^{-1}$V$^{-\alpha}$. We will further consider $U_{00} \equiv U_0$ and $\alpha=2$. The relation for $n$ can be expressed as 

\begin{equation}
\label{EqFitting4}
n = \lambda(E-E_0)(U_\mathrm{S}-U_0) \cdot \mathrm{ln} [a (U_\mathrm{S}-U_0)^2 i t]. 
\end{equation} 

The disadvantage of this equation is its divergence at $t=0$. By taking  the derivative of Eq.~\ref{EqFitting4}, we obtain the differential equation Eq.~\ref{EqDerivation}. By solving this equation, we obtain an integration constant: "+1" in the logarithm, see Eq.~\ref{EqFitting3}. This ensures the initial condition $n=0$ at $t=0$. With the integration constant known, we repeated all the fitting in Fig.~\ref{Fig10}a. It turns out that the impact on all values obtained from the fitting is very small, as the "+1" term in the logarithm is negligible for most data points. The only fitting constant that is affected significantly is $a$. The new value of $a=10$~A$^{-1}$s$^{-1}$V$^{-\alpha}$ was used in Fig.~\ref{Fig10}d.

We note that the fitting presented in this paper leads to the final equation \ref{EqDerivation} in the form d$n$/d$t \approx R \mathrm{e}^{-n}$. We have tried other ways of fitting the data, especially those leading to the Langmuir-type behavior d$n$/d$t \approx R (1-\theta)$. The experimental data do not fit such a behavior well.

\end{document}